# EFFICIENT SOUND CARD BASED EXPERIMENTION AT DIFFERENT LEVELS OF NATURAL SCIENCE EDUCATION

**Zoltan Gingl, Robert Mingesz and János Mellár,** *Department of Technical Informatics, University of Szeged*
**Balazs Lupsic and Katalin Kopasz,** *Department of Experimental Physics, University of Szeged*

## Abstract

Sound cards, which count as standard equipment in today's computers, can be turned into measurement tools, making experimentation very efficient and cheap. The chief difficulties to overcome are the lack of proper hardware interfacing and processing software. Sound-card experimentation becomes really viable only if we demonstrate how to connect different sensors to the sound card and provide suitable open-source software to support the experiments. In our talk, we shall present a few applications of sound cards in measurements: photogates, stopwatches and an example of temperature measurement and registration. We also provide the software for these applications.

## 1. Introduction

Physics and other natural science education can't be effective without properly designed, efficient, transparent and informative experiments. Using traditional instrumentation and experimental tools are important from the historical point of view, however most schools and universities run out of these, while modern measurement techniques should also play an important role and of course can be much more efficient.

Today's advanced, widely available and economic electronic solutions allow us to use sensors, digital equipments and personal computers to build wide variety of instruments and experimental setups, measure and display various physical quantities in real time, help students to understand more easily the physical phenomena and their description.

There are a broad range of computer controlled experimentation tools, data acquisition devices and displaying, analysing software on the market, but they are either too expensive or not flexible and efficient enough in most cases, probably can only be used for demonstration experiments. On the other hand, it would be very desirable to allow the students to make the experiments by themselves that needs many instances of the same experimental setup and instruments in the classroom or even at home, where the student could do experimental homework or find out some new experiments – this would mean much more room for creativity, motivation and efficient learning.

These problems are addressed by many researchers, there are several publications about simple and smart experiments, measurements that can be carried out by using multimedia computers by utilizing their built-in or their peripherals' digitizing and sensing components like the sound card, webcam, mouse and more. For example, a prepared computer mouse can be used as a stopwatch (Ganci 2009) or as a displacement recording device for a pendulum experiment (Gintautas et al 2009), a webcam can be used as a spectrograph (Grove et al 2007) or as a general measuring tool (Nedev et al 2006); the sound card can be used to measure the speed of sound (Martin 2001) or the time of flight of a free falling ball (Hunt et al 2002). Hobbyists also post their solutions on many web pages, show do-it-yourself (DIY) experiments.

Although many of these solutions are really useful and very cheap, the teacher should be very careful, because such instruments do not have guaranteed accuracy, they might teach some unusual, not recommended way, might be misleading, didactically poor or even unacceptable. Since the main goal is to keep the costs as low as possible, some freely available software is used for the measurements like in the case of sound card experiments, where the Audacity is quite popular among experimenters. Unfortunately these software are not designed for such experimentation, therefore they are typically very inconvenient, does not show the most important quantities in real time (for example, reading out the time between events needs considerable effort). Sometimes even the professional, dedicated software can be unhandy; we know that very minor difference can make a software much more popular than similar ones.

We can conclude that developing very simple, cheap and widely available experiments can be very useful, but careful instrumentation considerations and dedicated software development are needed.

In the following we'll give an overview about how to use multimedia computers' sound card – the only interface of the computer that can accept and generate voltages directly – to solve some instrumentation and experimentation problems in a very cheap and simple way. We have also developed various dedicated open-source software that can also be used as a reference development and basis for further modifications and upgrades.

## 2. Sound card photogates

Photogates are very popular devices that allow the teacher to measure mechanical events quite accurately. The photogate works like a light controlled switch; it is based on light detectors – like phototransistors or photoresistors – that conduct well if they are illuminated. Mechanical events can block the light beam from the detector, therefore they can be observed electronically.

The sound card can accept directly connected photodetectors if the computer runs the required signal processing software this way extremely cheap yet precise photogates can be built (Gingl et al 2011a).

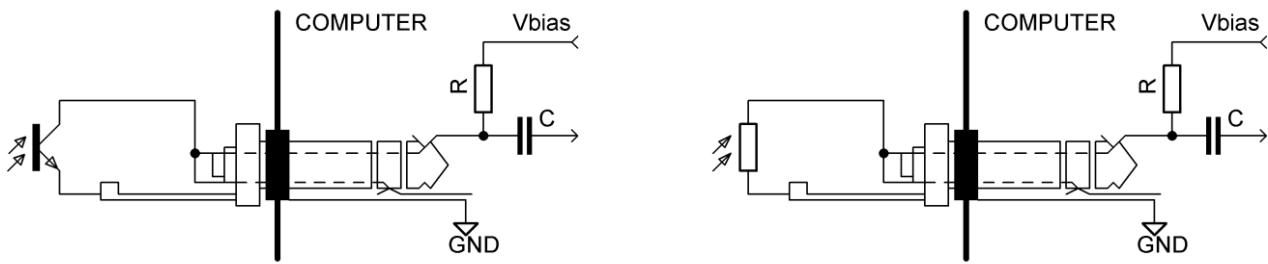

*Figure 1: Phototransistors and photoresistors can be directly connected to the microphone input. The photodetector and the sound card's internal resistor R and Vbias voltage forms a light-controlled voltage divider, therefore a voltage pulse can be measured, if the light beam is blocked from the detector.*

### 2.1. Pendulum experiment

Probably the most common experiment to demonstrate the performance of photogates and associated software is the pendulum experiment. The period of the oscillation is close to a second, while the sound card photogate provides resolution of better than one tenth of a millisecond. Our free photogate software (http://www.inf.u-szeged.hu/noise/edudev/Photogate/) provides real time display of the waveforms and also detects the light interruption time instants cause by the movement of the bob. If the diameter of the bob is entered, the software even calculates the approximate instantaneous speed and plots its history also in real time. This way the experiment and measurement principle is very transparent and the teacher can demonstrate that the period does not change significantly if the amplitude gets smaller while the decreasing speed is plotted, see Figure 2. Note that many other periodic movements can be easily monitored including the oscillation of a mass on a spring, rotation of a wheel or fan.

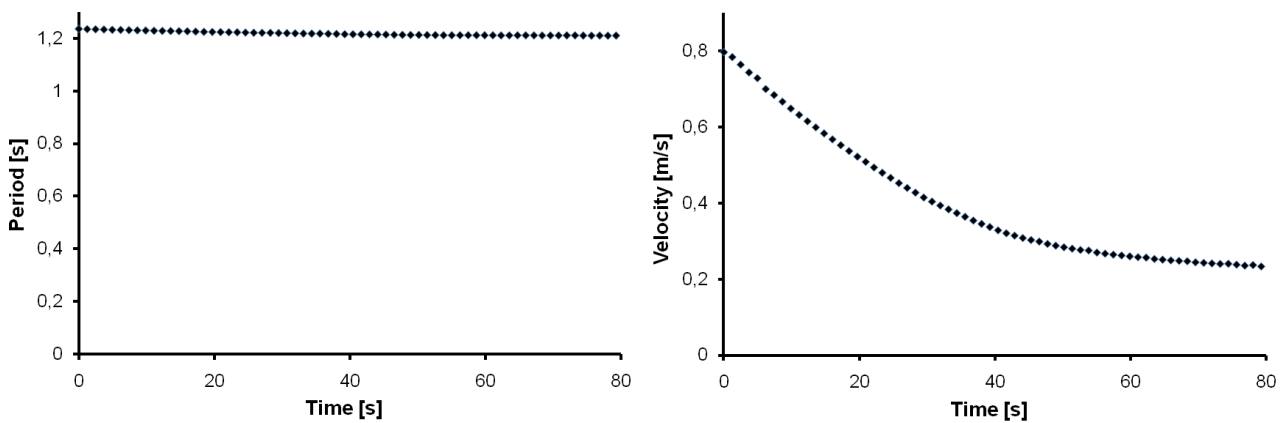

*Figure 2: Period and velocity of a pendulum as a function of time measured by the sound card photogate.*

### 2.2. Two free fall experiments

In the above it was shown how periodic movements can be investigated by the photogate using the associated software. The universal software allows monitoring of transient processes as well, the measured data are remembered and can be copied to any spreadsheet software by 2-3 mouse clicks. A picket fence is often used to investigate the free fall, only one photodetector is needed. In order to demonstrate the performance of sound card photogates, we have placed five strips of 19mm black tape on a transparent plastic ruler 5cm from each other, and dropped between the detector and light source. This way we could measure the displacement versus time and also could estimate the instantaneous velocity by measuring the time of light beam blocked by the 19mm tape.

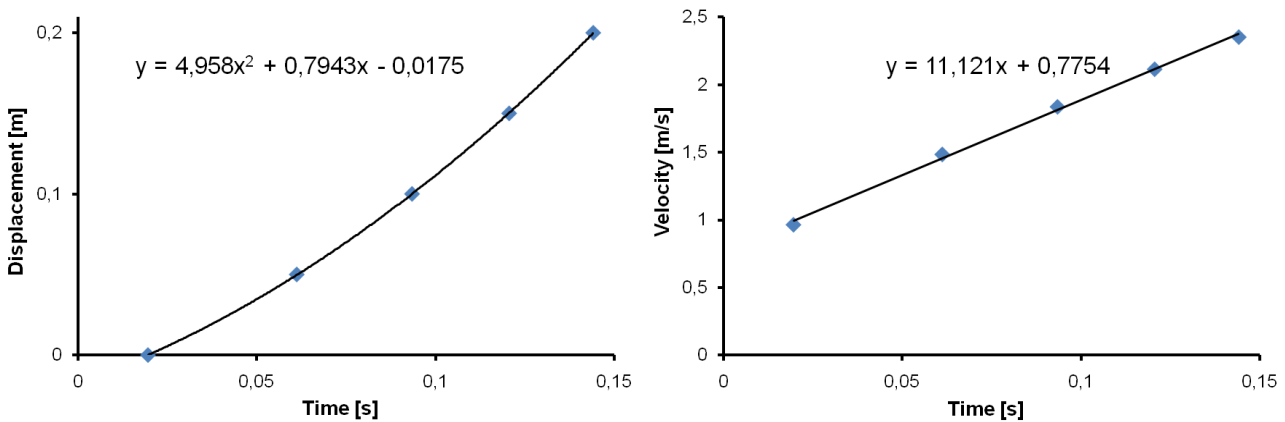

*Figure 3: Displacement and estimated velocity of a free falling picket fence. Fitting a parabola on the displacement gives g=9,916m/s², linear fit on the estimated velocity is less accurate, shows g=11,121m/s².*

Using multiple photodetectors provides additional flexibility, any object can be dropped. In this case we detect the light interruptions at least in two locations and measure the time of flight with a stopwatch-like method (Gingl et al 2011b). We have developed an open-source sound card stopwatch software (http://www.inf.u-szeged.hu/noise/edudev/Stopwatch/) to support this measurement mode. The measured accuracy is below 0,1ms, therefore it is even possible to show that a ping-pong ball needs longer time to fall than an iron ball.

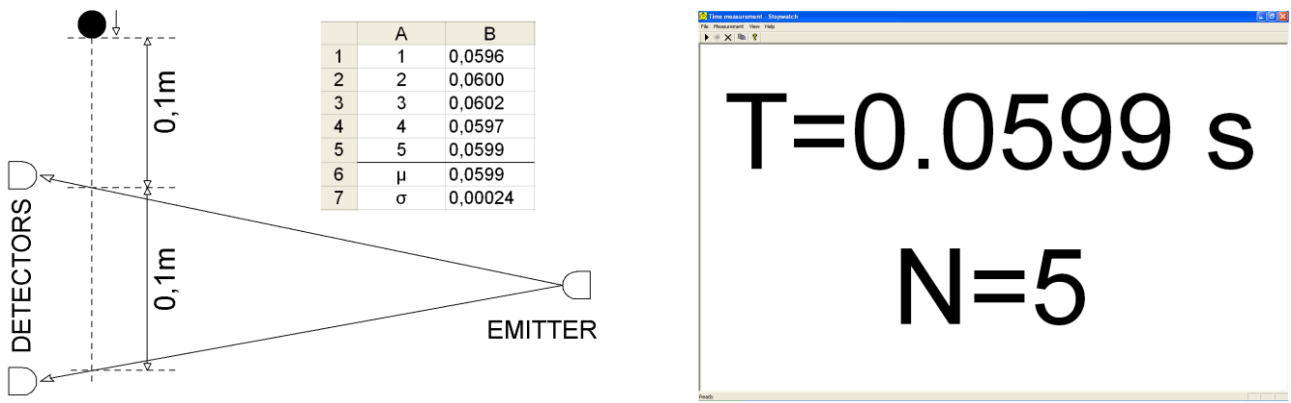

*Figure 4: Iron ball free fall experiment demonstrates the high precision of the sound card stopwatch. The five independent trials show very low uncertainty, the standard deviation is only 0,24ms, 0,4% of the mean.*

### 3. Sound card Ohmmeter and thermometer

In the above examples the sound card was only used to measure time that can be accurately done due to the rather precise time base of the sampling process. However, the sound card can also measure amplitude, but here the user must be very careful, only experienced teachers should use the sound card for such purposes. The sound card is designed for high audio signal fidelity that needs high linearity and high resolution. However, the absolute accuracy is poor and due to the automatic volume and tone adjustments of the hardware and operating system the measurement range can be changed unexpectedly. However, in some cases it is possible to use the sound card for amplitude measurements. The user must always do calibration before measurements, should be aware of other running software and sound card settings of the operating system – many things to keep in mind. Since the sound card has an output as well, this can be used to generate a sinusoidal signal that drives a voltage divider formed by a known and resistive sensor like a potentiometer, photoresistor or thermistor, and using stereo mode impedance can also be measured (Klaper et al 2008). This way the sound card can be used to measure resistance and many other physical quantities that have an associated resistive sensor. The measurement range is limited to 1k-100k to keep calibrated accuracy of about 1-2%.

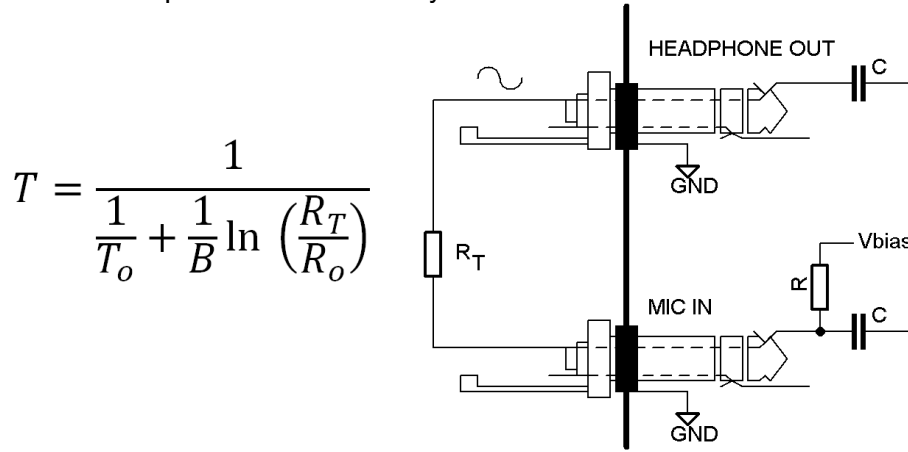

$$T = \cfrac{1}{\cfrac{1}{T_o} + \cfrac{1}{B} \ln\left(\cfrac{R_T}{R_o}\right)}$$

*Figure 5: A thermistor $R_T$ is connected between the headphone output and microphone input. R and $R_T$ forms a voltage divider for the sinusoidal headphone output voltage, therefore $R_T$ can be measured. The formula on the left hand side can be used to estimate the temperature in Kelvins, where $T_0$ is 293K, $R_0$ is the resistance of the thermistor at temperature $T_0$ and B can be obtained from the datasheet.*

We have written an open-source software in Java (Gingl et al 2011c) that can be used as a simple Ohmmeter, as a thermometer or even as a chart recorder in Windows or Linux environments. Depending on the user's choice, the recorded data can be displayed in the units of Ohms or in Celsius/Fahrenheit, if a thermistor is connected and its parameters ($R_0$ and B, see the Figure 5) are entered. The application can export the displayed plot into a jpeg file, can also export the data into a text file using the specified column separators. What is even more handy, the data can be copied to the clipboard, and the user can paste it into his favorite spreadsheet application. Calibration is simple and requires only two steps. First the user is asked to connect the two measuring terminals, this represents zero resistance. The user should change the playing and/or recording volume as necessary to prevent overdriving the input or getting too small signal. In the next step a 10k resistor with at least 1% accuracy should be connected. After doing so, the system is calibrated and ready for the measurement. The performance of the sound card thermometer can be demonstrated by a cooling experiment, Figure 6 shows the measurement result.

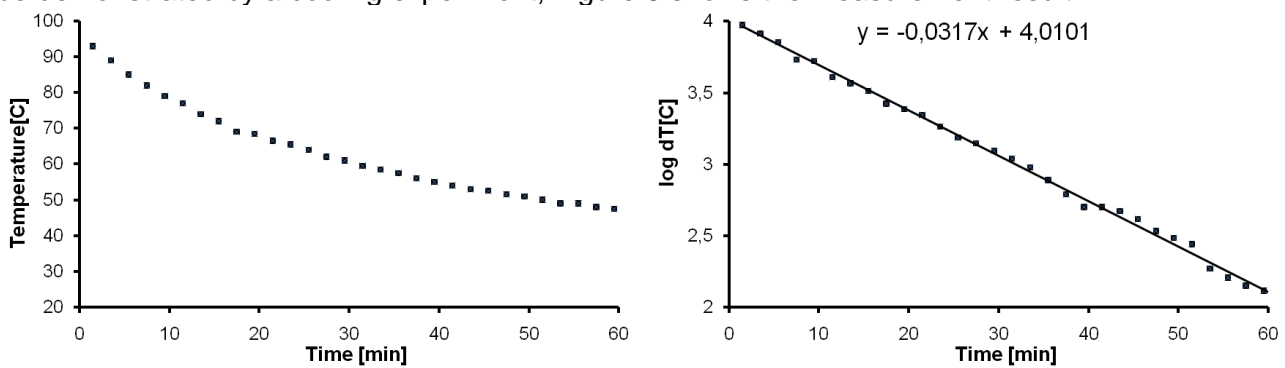

*Figure 6: Cooling process of a glass of water measured with the sound card thermometer. The plot on the right hand side shows the natural logarithm of the temperature difference from the final temperature as a function of time. The time constant is 1/0,0317min=31,55min.*

Note, that one may want to provide additional protection of the audio input/output ports by inserting series resistors of 1-2 kOhms at both connectors; alternatively cheap USB sound cards can also be used to avoid any possible damage to the internal audio circuitry.

## 4. Conclusion

All multimedia computers have a built-in sound card – the only input/output interface that can directly accept and generate analogue signals, voltages. Although the sound card has limited capabilities (can only handle alternating signals in a limited frequency and range) and it is not designed to be used as an instrument, it can efficiently help experimental education at a very low price. Several sensors can be directly connected to the sound card input including photoresistors, phototransistors, thermistors, if proper signal processing is provided by the accompanying software.

In this paper we have shown photogate and stopwatch examples for precision real-time investigation of mechanical movements and we have also demonstrated that a single thermistor connected to the sound card can provide a rather accurate thermometer, temperature chart recorder. Many sound card experiments are based on commercially available audio software like Audacity, but unfortunately these do not provide convenient and efficient signal processing therefore we have also developed dedicated open-source software to support transparent, efficient experimentation and data presentation.

The available demonstrations and examples, very low price, ease-of-use allow the teachers and students to turn their computers into measurement devices, students can even make experiments as a part of their homework, use their creativity to improve the experiments or they can even modify the open-source software to their needs, learn data visualization and signal processing programming looking at the source code.

Note that experienced users can build simple very low cost remote controlled experiments based on the sound card input/output. The AC coupling of the sound card inputs and outputs supports several experiments, in other cases most likely voltage-to-frequency and frequency-to-voltage converters will be used to extend the possibilities with generation and measurement of DC voltages.

For more information and software download please visit http://www.inf.u-szeged.hu/noise/edudev.